\documentclass[preprint,onecolumn,nofootinbib]{revtex4}
\pdfoutput=1
\usepackage[colorlinks=true,linkcolor=blue,urlcolor=blue,filecolor=black,citecolor=red,pdfstartview=FitV,pdftitle={},pdfsubject={},pdfkeywords={},pdfpagemode=None,bookmarksopen=true]{hyperref}
\usepackage{graphicx}
\usepackage{amsmath}
\usepackage{amsfonts}
\usepackage{amssymb,ulem}
\usepackage{color}%
\usepackage{CJK}
\usepackage{dcolumn}
\usepackage{setspace}

\newcommand{\f}{\begin{equation}}
\newcommand{\ff}{\end{equation}}
\newcommand{\fa}{\begin{eqnarray}}
\newcommand{\ffa}{\end{eqnarray}}

\begin{document}

\title{{Linear Instability of the Charged Massless Scalar Perturbation in Regularized 4D Charged Einstein-Gauss-Bonnet Anti de-Sitter Black Hole}}
\author{Peng Liu}
\email{phylp@email.jnu.edu.cn}
\author{Chao Niu}
\email{niuchaophy@gmail.com}
\author{Cheng-Yong Zhang}
\email{zhangcy@email.jnu.edu.cn, corresponding author}
\affiliation{
  $^{1}$ Department of Physics and Siyuan Laboratory, Jinan University, Guangzhou 510632, P.R. China
}
\begin{abstract}
  {We study the linear instability of the charged massless scalar perturbation in regularized 4D charged Einstein-Gauss-Bonnet-AdS black holes by exploring the quasinormal modes}. We find that the linear instability is triggered by superradiance. The {charged massless scalar perturbation} becomes more unstable when increasing the Gauss-Bonnet coupling constant or the black hole charge. Meanwhile, {decreasing} the AdS radius will make the {charged massless scalar perturbation} more stable. The 
  stable region in parameter space $(\alpha,Q,\Lambda)$ is given. Moreover, we find that the {charged massless scalar perturbation} is more unstable for larger scalar charge. The modes of multipoles are more stable than that of the monopole.
\end{abstract}
\maketitle
\tableofcontents

\section{Introduction}

The perturbations of black holes are powerful probes to disclose the stability of the black holes and have been studied intensively for decades. The linear (in)stability of the black hole can be characterized by the quasi-normal modes (QNMs). 
If the imaginary part of the {QNMs} are positive, the perturbation amplitude will grow exponentially and implies instability. The QNMs provide the finger-prints of black holes and are related to the gravitational wave observations \cite{Konoplya2011}. In four-dimensional spacetime, the black holes, such as the Schwarzschild black holes, Reissner--Nordstr\"om (RN) black holes and Kerr black holes, are stable under neutral scalar, electromagnetic field or gravitational perturbations in general \cite{Konoplya2011}, regardless of whether the black holes are in asymptotic flat, de Sitter (dS) or anti-dS (AdS) spacetimes\footnote{The case for Kerr-Newman spacetime is subtle due to the difficulty of decoupling the variables. Numerical works strongly support that they are stable \cite{Pani2013}. But this problem has not been settled.}. However, the four-dimensional black holes can be unstable provided superradiance occurs \cite{Brito2015,Zhu2014,Zhang2014}. The perturbations in higher dimensional spacetimes and alternative theories of gravity have also attracted a lot of attentions \cite{Li2019RNdS}.

Recently, a regularized four-dimensional Einstein-Gauss-Bonnet (EGB) gravity theory was proposed \cite{Glavan20192019a}, in which the Gauss-Bonnet coupling constant was rescaled with $\alpha\to\alpha/(D-4)$ in the limit $D\to4$ and  novel black hole solutions were found \cite{Fernandes2003,4DEGBSolutions}. This has stimulated a lot of studies, as well as doubts \cite{Kobayashi2004,Aoki2020,4DStability,Zhang2020Super,Liu2020,4DEGBShadow,4DEGBThermo,4DEGBOthers}. Fortunately, some proposals have been raised to circumvent the   issues of the regularized 4D EGB gravity,
including adding an extra degree of freedom to the theory \cite{Kobayashi2004}, or breaking the temporal diffeomorphism invariance \cite{Aoki2020}, where a well-defined theory can be formulated.
The regularized black hole solutions can still be derived from these more rigorous routes. Thus it is worth to study the perturbations of these black hole solutions. In fact, the stability of the regularized four-dimensional black hole has been studied from many aspects \cite{4DStability}, of which most focused only on the neutral cases. We studied the charged massless scalar perturbations {in} the regularized 4D charged EGB black holes {with} asymptotic flat and dS spacetimes in \cite{Zhang2020Super,Liu2020}, respectively. It was found that the {charged massless scalar perturbations} in asymptotic flat spacetime
are always stable, while those in asymptotic dS spacetime suffer from a new kind of test field instability where not all the modes satisfying the superradiant condition are unstable.

Since the boundary conditions in AdS spacetime are different from those in asymptotically flat or dS spacetimes, it can be expected that the behaviors of perturbations on the black holes in AdS spacetime would be different from those in flat or dS spacetime.
Thus we study the charged scalar perturbation {in} the regularized 4D charged black hole in AdS spacetime in this paper. We find that the asymptotic iteration method used in \cite{Zhang2020Super,Liu2020} does not work well here. Instead, we adopt another numerical method to calculate the frequencies of the perturbations.
The QNMs are worked out as a generalized eigenvalue problem and the results are checked by the time-evolution method.
We find that the {charged massless scalar} suffers from linear instability. However, unlike the dS case, all unstable modes here satisfy the superradiant condition.
The effects of the Gauss-Bonnet coupling constant, the black hole charge, the scalar charge and cosmological constant were analyzed in detail.

This paper is organized as follows. In section \ref{4degb}, we describe the regularized 4D charged EGB black hole in AdS spacetime and the parameter region allowing an event horizon of the black hole. In section \ref{qnm_num_method} we elaborate the method to calculate the quasinormal modes of the {charged scalar perturbation in} regularized charged black hole. In sections \ref{results}, we study the effects of the Gauss-Bonnet coupling constant, the black hole charge, the scalar field charge and the cosmological constant on the QNMs in detail. Section \ref{sec:discussion} is our discussion.

\section{4D Einstein-Gauss-Bonnet gravity}\label{4degb}

In spherically symmetric spacetime, the electrovacuum solution of the four-dimensional EGB gravity in AdS spacetime is given by \cite{Fernandes2003}
\begin{equation}\label{eq:metric}
  ds^{2}=-f(r)dt^{2}+\frac{1}{f(r)}dr^{2}+r^{2}(d\theta^{2}+\sin^{2}\theta d\phi^{2}),
\end{equation}
where the metric function
\begin{equation}\label{eq:fexp}
  f(r)=1+\frac{r^{2}}{2\alpha}\left(1-\sqrt{1+4\alpha\left(\frac{M}{r^{3}}-\frac{Q^{2}}{r^{4}}{+}\frac{\Lambda}{3}\right)}\right),
\end{equation}
and the gauge potential
\begin{equation}
  A=-\frac{Q}{r}dt.
\end{equation}
Here, $M,\,Q$ are the mass and charge of the black hole, respectively. In \eqref{eq:fexp} the cosmological constant is {$\Lambda=-3/l^2$}, where $l$ is the cosmological radius. As $r\to\infty$, the solution approaches an asymptotic AdS spacetime if
 \begin{equation}\label{eq:adsreq}
  \frac{1}{2\alpha} \left( 1-\sqrt{1{+}\frac{4\alpha\Lambda}{3}} \right)>0.
\end{equation}
This condition can be satisfied only for {negative} $\Lambda$ no matter $\alpha$ is positive or negative. When $\alpha\to0$, the solution goes back to the RN-AdS black hole. Note that the solution (\ref{eq:metric})
coincides formally with {those} obtained from conformal anomaly and quantum corrections \cite{Cai20019} and those from Horndeski theory \cite{Kobayashi2004}.

\begin{figure}[htbp]
  \centering
  \includegraphics[width = 0.5\textwidth]{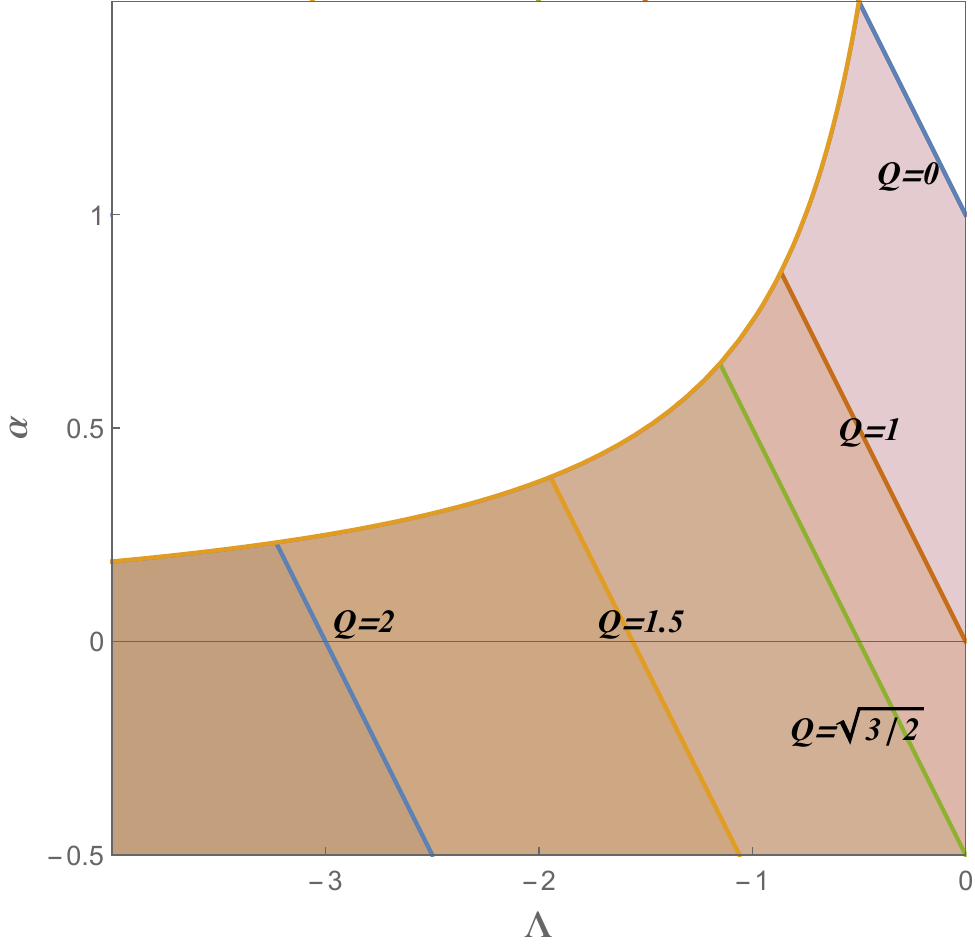}
  \caption{The parameter region that allows the existence of a black hole with metric function (\ref{eq:fexp}). The allowed region for a larger $Q$ is a subset of that for a smaller $Q$. }
  \label{fig:allowedregion}
\end{figure}

We fix the black hole event horizon $r_{+}=1$ in this paper. To ensure $f(r_+)=0$, there should be $\alpha>-1/2$. The mass $M$ can be expressed as
\begin{equation}
  M=\frac{1}{3} \left(3 \alpha {-}\Lambda +3 Q^2+3\right).
\end{equation}
The parameter region that allows the existence of a black hole with positive Hawking temperature on the event horizon is determined by $f'(r_+)>0$. This leads to an inequality which is very similar to the  case in dS spacetime \cite{Liu2020}.
\begin{equation}\label{eq:allowed}
  Q^2 + \alpha {+} \Lambda <1.
\end{equation}
The region plot of the allowed parameter region is given in Fig. \ref{fig:allowedregion},
from which we can find that the allowed region {(shaded region)} shrinks when increasing $Q$. However, unlike  the case in dS spacetime, the black hole charge $Q$ is unbounded in asymptotic AdS spacetime here.

\section{Quasinormal modes and numerical methods}\label{qnm_num_method}

We study the linear stability of {a massless charged scalar perturbation $\psi$ in} the black hole solution (\ref{eq:metric}). It is known that fluctuations of order $O(\epsilon)$ in the scalar field in a given background induce changes in the spacetime geometry of order $O(\epsilon^2)$ \cite{Brito2015}.
To leading order, we can study the perturbations on the fixed background geometry that satisfies
\begin{equation}\label{eq:eom1}
  D_\mu D^\mu \psi =0,
\end{equation}
where $D_\mu \equiv \nabla_\mu - ieA_\mu$, and $e$ is the charge of the test scalar field.

It is more convenient to work under the ingoing Eddington-Finkelstein coordinate
\begin{equation}\label{eq:efcoord}
  v=t + r_*,
\end{equation}
 when studying the time evolution of the perturbation. Here $r_*$ is the tortoise coordinate defined with $dr_* = dr/f$. In this coordinate, the line element becomes
\begin{equation}\label{eq:metric2}
  ds^2 = -f dv^2 + 2 dvdr + r^2 d \Omega^2_2.
\end{equation}
The Maxwell field in Eddington-Finkelstein coordinates is
\begin{equation}\label{eq:em2}
  A = -\frac{Q}{r} dv,
\end{equation}
where we get rid of the {spatial} component of the gauge field by a gauge transformation.

The equation of motion \eqref{eq:eom1} is separable by taking the following form
\begin{equation}\label{eq:psiform}
  \psi = \frac{\phi(v,r)Y_{lm}(\theta,\varphi)}{r},
\end{equation}
{where $Y_{lm}$ is the spherical harmonic function.}
Inserting \eqref{eq:psiform} into \eqref{eq:eom1} we have
\begin{equation}\label{eq:main1}
  f\partial_r^2 \phi + f' \partial_r \phi - 2ie A_v \partial_r \phi - ie A'_v \phi + 2\partial_v \partial_r \phi + V(r)\phi =0,
\end{equation}
where the prime $'$ denotes the derivative with respect to $r$, and the effective potential is
\begin{equation}\label{eq:effpotential}
  V(r) = -f'(r) /r - l(l+1) /r^2.
\end{equation}
\begin{figure}[htbp]
  \centering
  \includegraphics[width = 0.6\textwidth]{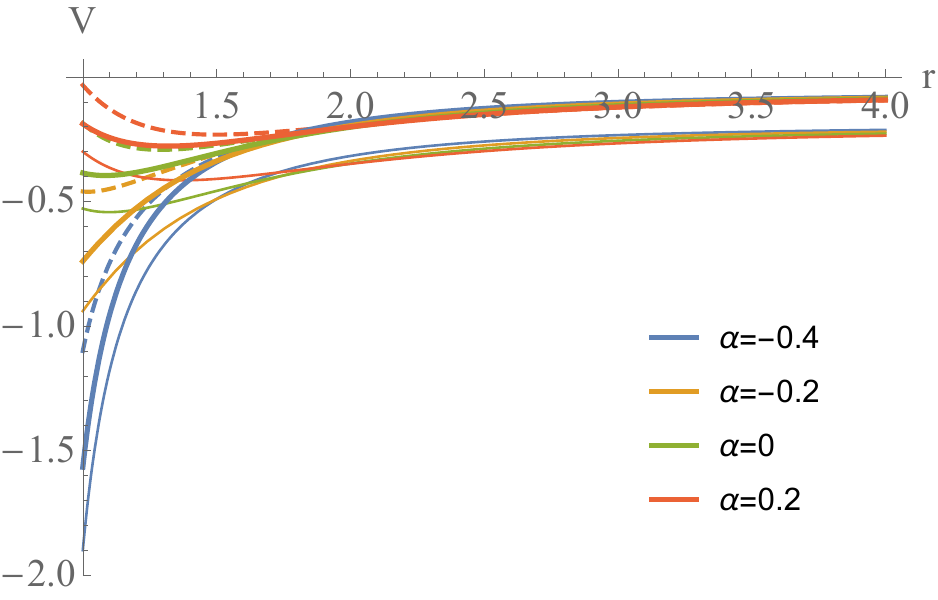}
  \caption{The effective potential when $l=0$. The thick lines for $Q=0.6,\Lambda={-}0.1$, the thin lines for $Q=0.6,\Lambda={-}0.3$, the dashed lines for $Q=0.8,\Lambda={-}0.1$.}
  \label{fig:effpotential}
\end{figure}
We show the effective potential when $l=0 $ in Fig. \ref{fig:effpotential}. When $Q$ and $\Lambda$ are fixed, the effective potential well becomes shallower as $\alpha$ increases. When $\Lambda$ and $\alpha$ are fixed (the thick lines and dashed lines), the effective potential well becomes shallower as $Q$ increases. When $Q$ and $\alpha$ are fixed (the thick lines and thin lines), the effective potential becomes deeper as $\Lambda$ increases. We will see that these behaviors of effective potential are related to the linear instability structure of the {massless charged scalar perturbation on the black hole}.

In order to implement the frequency analysis on \eqref{eq:main1}, we {consider the following mode}
\begin{equation}\label{eq:phiomegav}
  \phi(v,r) = \phi(r) e^{-i\omega v}.
\end{equation}
Then equation \eqref{eq:main1} becomes
\begin{equation}\label{eq:main2}
  f\partial_r^2 \phi + f' \partial_r \phi - 2ie A_v \partial_r \phi - ie A'_v \phi -2i\omega \partial_r \phi + V(r)\phi =0.
\end{equation}
A {more} convenient approach to implement the numerics is to work under $ z \equiv r_+ / r $ coordinate, such that we can solve the equation of motion in a bounded region $ z \in [0,1] $. This coordinate has been widely adopted in solving gravitational background solutions and perturbations \cite{Lingbefore}. Here the ingoing condition at horizon is satisfied by setting the time-dependent factor $e^{-i \omega v}$ {and requiring a regular $\phi(r)$ near the horizon}. {The asymptotic AdS requires a vanishing $\phi(r)$ at the boundary, that can be realized by extracting $1/r$ from $\phi$.}

The next step is to find the QNMs, i.e. the modes with complex {frequency} $ \omega  = \omega_{R} + i \omega_{I}$. When the imaginary part $ \omega_I >0 $, the perturbation amplitude grows exponentially with time. When the amplitude becomes large enough, the back reaction of the perturbation on the geometry can not be ignored and implies that the system may become unstable. However, at this stage, the linear approximation adopted in this paper is inadequate and the full nonlinear studies are required.

The radial equation \eqref{eq:main2} is generally hard to
solve analytically, except for a few instances such as the pure (A)dS spacetimes and Nariai spacetime \cite{Konoplya2011}.
Various numerical methods were developed to obtain the QNMs, such as the WKB method, perturbation method, and iteration methods \cite{qnmmethods}.

Recently, a new method that re-casts the search for QNMs to a generalized eigenvalue problem by discretizing the equation of motion has been proposed \cite{newmethod}.
In this paper, we discretize the spatial coordinate with Gauss-Lobatto grid
with $N$ collocation points. It has been shown that the Gauss-Lobatto grid is an efficient method for solving the eigenvalue problems \cite{Boyd:2001}.
We  examined this in our calculations and found that the Gauss-Lobatto collocation method requires significantly fewer points than the uniform collocation method to achieve the same accuracy.
With the discretization, we arrive at a discretized version of  \eqref{eq:main2},
\begin{equation}\label{eq:matrix}
  \left(\mathcal{M}_0 + \omega \mathcal M _1 \right) \vec \phi = 0,
\end{equation}
where $\mathcal M_{0,1}$ are  {$N\times N$} complex-valued matrices and independent of $\omega$.
The eigenvalues $\omega$ in \eqref{eq:matrix} can be efficiently solved by {\texttt{Eigenvalues[-$\mathcal M_0$,$\mathcal M_1$]}} with {\texttt{Mathematica}}. Equivalently, one may also solve the roots of $\det \left(\mathcal{M}_0 + \omega \mathcal M _1 \right) =0 $ to find the eigenvalues.
Resultantly, a finite set $\{\omega\}$ with $N$ elements can be obtained, of which some are spurious QNMs. The results are reliable only if they are convergent when increasing the collocation density. The fundamental mode is the one with the largest imaginary part.

\section{Results}\label{results}

We study the linear instability of the {massless charged scalar field in} 4D EGB model by showing the relation between the dominant QNMs and the system parameters. We first explore the effect of  $\alpha, \, Q,\, \Lambda$ and $e$ on the QNMs, respectively. After that we study the comprehensive linear instability structure of the {massless charged scalar perturbation} by showing the stable region in the allowed parameter space.

\subsection{QNMs vs $\alpha$}

In the asymptotic dS case, the effect of $\alpha$ on the linear instability depends on the specific  {values} of $Q$ and $\Lambda$ \cite{Liu2020}. It is especially important to find out the role the GB coupling constant $\alpha$ plays in the linear instability structure of AdS case here.

\begin{figure}[htbp]
  \centering
  \includegraphics[width = 0.45\textwidth]{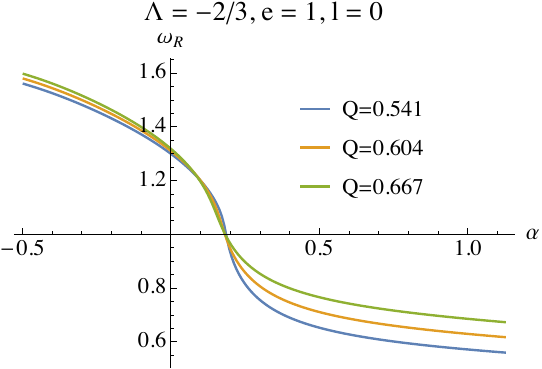}
  \includegraphics[width = 0.45\textwidth]{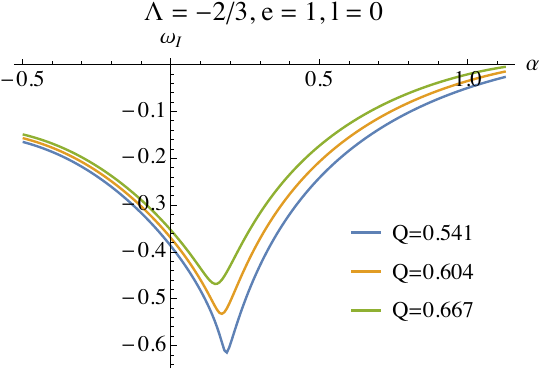}
  \caption{The left and the right plots are $\omega_R$ and $\omega_I$ of the  fundamental modes vs $\alpha$ at $\Lambda = {-}2/3,\,e=1,\,l=0$, respectively.}
  \label{fig:alongalpha1}
\end{figure}

As an example, we fix $\Lambda ={-} 2/3,\,e=1,\,l=0$ and demonstrate the QNMs vs $\alpha$ in Fig. \ref{fig:alongalpha1}. We can see that the real part of the fundamental mode $\omega_R$ decreases with $\alpha$ monotonically. By comparing the curves of different charges $Q$ we find that $\omega_R$ monotonically decreases with $Q$. Meanwhile, we can find that the imaginary part of the fundamental mode $\omega_I$ first decreases with $\alpha$ and then increases with $\alpha$. The imaginary part $\omega_I$ increases with $Q$, indicating that increasing $Q$ may lead to linear instability. However, in Fig. \ref{fig:alongalpha1} where we have fixed $e=1$, the $\omega_I$ is always negative and  no linear instability occurs.

\begin{figure}[htbp]
  \centering
  \includegraphics[width = 0.45\textwidth]{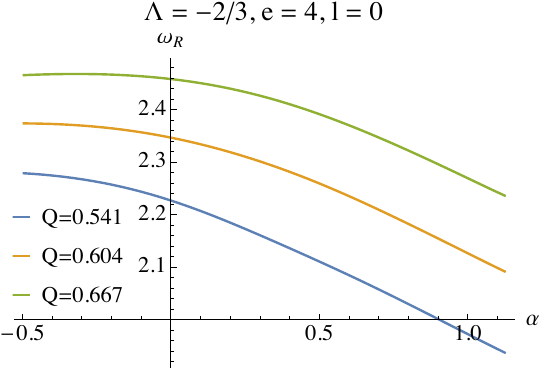}
  \includegraphics[width = 0.45\textwidth]{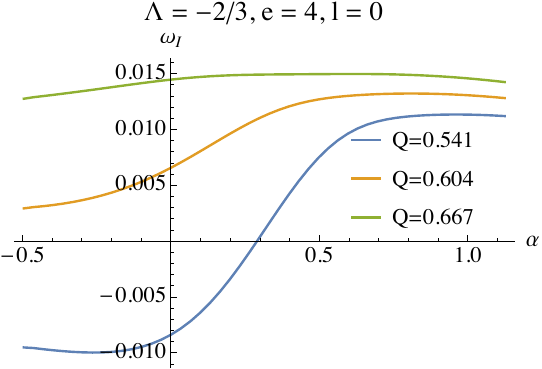}
  \caption{The left and the right plots are $\omega_R$ and $\omega_I$ vs $\alpha$ of the  fundamental modes at $\Lambda ={-} 2/3,\,e=4,\,l=0$, respectively.}
  \label{fig:alongalpha2}
\end{figure}
We show the fundamental modes vs $\alpha$ at larger scalar charge $e=4$ in Fig. \ref{fig:alongalpha2}. Apparently, the $\omega_R$ also monotonically decreases with $\alpha$, which is in accordance with that of the case when $e=1$. For the imaginary part, we find that when $Q$ is relatively small, $\omega_I$ first decreases with $\alpha$   and then increases with $\alpha$ (see the blue curve in the right plot of Fig. \ref{fig:alongalpha2}), this is similar to that of $e=1$. However, when $\alpha$ is relatively large, $\omega_I$ can become positive and {lead} to linear instability here. Therefore the linear instability occurs only at large enough value of perturbation charge $e$. For intermediate values of $Q$, we find that $\omega_I$ can {increase} with $\alpha$ monotonically. For large values of $Q$, the $\omega_I$ can decrease with $\alpha$ for relatively large $\alpha$. In addition to that, we can see that both $\omega_R$ and $\omega_I$ increases with $Q$ for $e=4$. The more comprehensive dependence of the linear instability on the scalar charge $e$ is given in subsection \ref{subsec:vse}.

From the viewpoint of the effective potential shown in Fig. \ref{fig:effpotential}, increasing  $\alpha$ makes the effective potential well shallower. An effective potential barrier appears near the horizon when $\alpha$ is large enough. The scalar perturbation is harder to be absorbed by the black hole and can accumulate in the effective potential well. As a consequence, {the massless charged scalar field} tends to be unstable as $\alpha$ increases and the imaginary part of the frequency $\omega_I$ tends to be positive.

\begin{figure}[htbp]
  \centering
  \includegraphics[width = 0.45\textwidth]{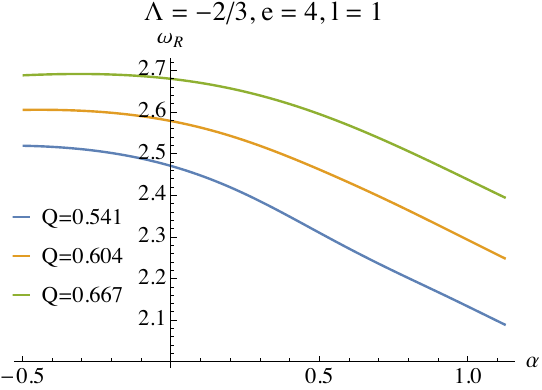}
  \includegraphics[width = 0.45\textwidth]{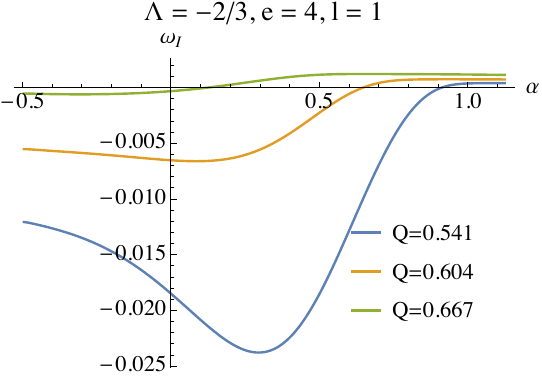}
  \caption{Fundamental modes vs $\alpha$ at $\Lambda = {-}2/3,\,e=4,\,l=1$, where the left and the right plots are the real part and imaginary part of the fundamental modes, respectively.}
  \label{fig:ellno01}
\end{figure}

It has been shown modes of $l>0$ are usually more stable than those of $l=0$ \cite{Liu2020}. In present model, similar phenomenon can also be observed. We show the fundamental {modes} vs $\alpha$ at $e=4,\,l=1$ in Fig. \ref{fig:ellno01}. Comparing Fig. \ref{fig:alongalpha2} and Fig. \ref{fig:ellno01} we see that $\omega_I$ for $l=1$  is smaller than that of $l=0$, which implies that $l=1$ {perturbation} mode is indeed more stable than that of $l=0$ mode. For the real part of the fundamental {mode,} we find that $\omega_R$ for $l=1$ is larger than that of $l=0$, which corresponds to a more rapidly oscillating mode.

In this subsection, we find that increasing the coupling constant $\alpha$ can lead to linear instability, and this instability occurs only at large enough perturbation charge $e$. Moreover, the stability is enhanced when increasing the degree $l$.

\subsection{QNMs vs $Q$}\label{subsec:vsq}

\begin{figure}[htbp]
  \centering
  \includegraphics[width = 0.45\textwidth]{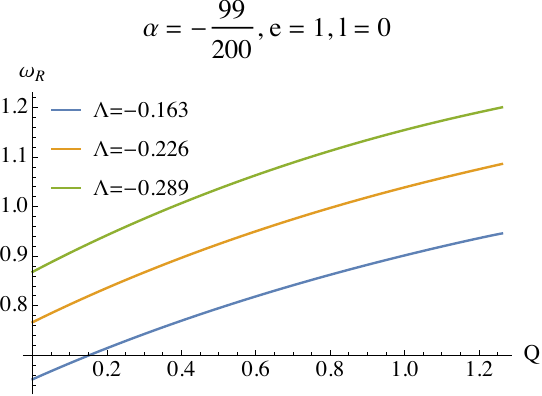}
  \includegraphics[width = 0.45\textwidth]{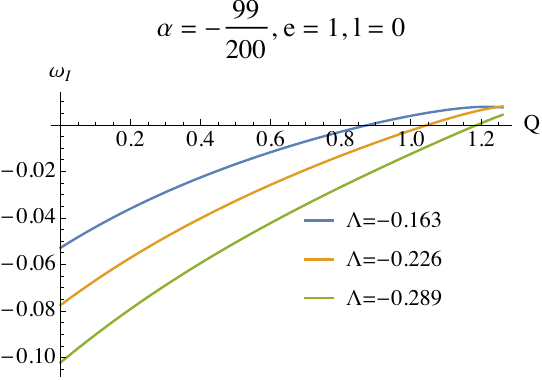}
  \caption{Fundamental modes vs $Q$ at $\alpha = -99/200,\,e=1,\,l=0$, where the left and the right plots are the real part and imaginary part of the fundamental modes, respectively.}
  \label{fig:forq1}
\end{figure}

In {the} previous subsection, we find that $\omega_I$ increases with $Q$ by comparing several examples at different $Q$'s (Fig. \ref{fig:alongalpha1} and Fig. \ref{fig:alongalpha2}). Here we explore more comprehensive relationship between the QNMs and the black hole charge $Q$.
In Fig. \ref{fig:forq1} we show the dependence of  QNMs on $Q$ when $\alpha = - \frac{99}{200},\,e=1,\,l=0$ as an example.
We  find that both $\omega_R$ and $\omega_I$ increase with $Q$. Especially,  $\omega_I$ can become positive when increasing $Q$, this means that increasing $Q$ will render the system more unstable and more oscillating. This is in accordance with the observations in the previous subsection.

Comparing the thick lines and dashed lines in Fig. \ref{fig:effpotential}, we see that increasing the black hole charge makes the effective potential at the horizon larger.  It is harder to absorb the perturbations by black holes with larger $Q$. The perturbations can be trapped more easily in the effective potential well. Thus {increasing} $Q$ makes the system more  unstable.

Also, by comparing the data at different values of $\Lambda$ we find that $\omega_R$ {decreases when increasing} $\Lambda$, while {$\omega_I$} {increases when increasing} $\Lambda$. This phenomenon suggests that {decreasing} $\Lambda$ will make the {charged massless scalar perturbation} more stable and less oscillating.

\subsection{QNMs vs $\Lambda$}
\begin{figure}[htbp]
  \centering
  \includegraphics[width = 0.45\textwidth]{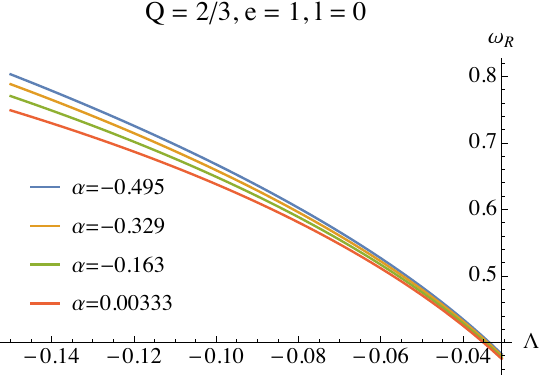}
  \includegraphics[width = 0.45\textwidth]{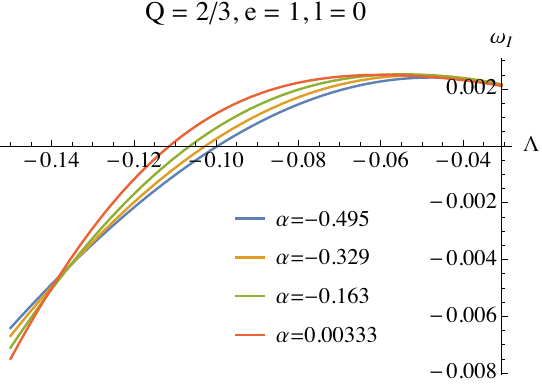}
  \caption{Fundamental modes vs $\Lambda$ at $Q = 1/10,\,e=1,\,l=0$, where the left and the right plots are the real part and imaginary part of the fundamental modes, respectively.}
  \label{fig:forlambda1}
\end{figure}

We show the QNMs vs $\Lambda$ when $Q = 1/10,\,e=1,\,l=0$ in Fig. \ref{fig:forlambda1}. From the left plot we see that $\omega_R$ monotonically {decreases} with $\Lambda$. For the imaginary part, when {$|\Lambda|$} is relatively small, $\omega_I$ is positive and {decreases} with $\Lambda$, the system is always unstable \footnote{We would like to mention that for substantially small values of {$|\Lambda|$}, our numerics cannot obtain stable results of the QNMs. Therefore, we only show $\Lambda {<-} 0.03$ where our numerics are precise enough.}. However, when $\Lambda$ {further decreases}, $\omega_I$ reaches its local maximum and then starts to decrease. When {decreasing} the $\Lambda$ {even} further, $\omega_I$ becomes negative and the system becomes stable. This again certifies {the observations in the last subsection} that the system becomes more stable {and more rapidly oscillating} when {decreasing} $\Lambda$.
From the potential behaviors in Fig. \ref{fig:effpotential}, we see that {decreasing} $\Lambda$ makes effective potential well deeper. The perturbations can be trapped more easily by black holes with larger $\Lambda$. Thus {decreasing} $\Lambda$ makes the system more stable.
Observing the  points where $\omega_I = 0$ we can also find that the system becomes more stable when increasing $\alpha$.

\subsection{QNMs vs $e$}\label{subsec:vse}

\begin{figure}
  \centering
  \includegraphics[width=0.7\textwidth]{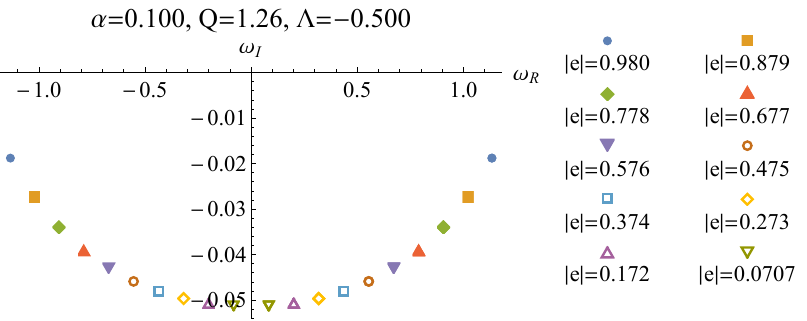}
  \caption{The fundamental modes for different values of $e$. Each color or marker corresponds to a pair of $e$'s with an absolute value $|e|$, where the left hand side is $e=-|e|$ and the right hand side $e=|e|$.}
  \label{fig:edependence}
\end{figure}

In this subsection, we systematically study the relationship between the linear instability {of the charged massless scalar field} and the scalar charge $e$. In principle, the scalar charge $e$ can be either positive or negative. However, here the linear instability structure with negative charge $-|e|$ is the same as that of the positive charge $|e|$.
Inserting the Maxwell field \eqref{eq:em2} into  \eqref{eq:main1}, we have
\begin{equation}\label{eq:main11}
  f\partial_r^2 \phi + f' \partial_r \phi - 2ieQ \partial_r \phi/r - ieQ \phi/r^2 + 2\partial_v \partial_r \phi + V(r)\phi =0,
\end{equation}
from which we  find that $e$ always appears in pairs with $iQ$, and the other terms  contain only $Q^2$. Accordingly, by separating the real part and imaginary part, the discrete eigenvalue equation \eqref{eq:matrix} can be decomposed into the following form
\begin{equation}\label{eq:diseig}
  \left[ ({\mathcal M}^R_0 + i\omega_I {\mathcal M}_1 ) + i ({\mathcal M}^I_0 - i\omega_R {\mathcal M}_1)\right]\vec \phi = 0.
\end{equation}
Here ${\mathcal M}_1$ is a pure imaginary matrix because $\omega$ always comes in pairs with $i$ in \eqref{eq:main2} and $\partial_r$ only contributes a real-valued differential matrix. The conjugate of \eqref{eq:diseig} becomes,
\begin{equation}\label{eq:diseigconj}
  \left[ ({\mathcal M}^R_0 + i\omega_I {\mathcal M}_1 ) + i (-{\mathcal M}^I_0 + i\omega_R {\mathcal M}_1)\right]\vec \phi^{*} = 0.
\end{equation}
Consequently, performing ${\mathcal M}^I_0 \to -{\mathcal M}^I_0$, which can result from performing $e\to -e$ since $\mathcal M^I_0$ is linear in $e$, we obtain a root $-\omega^{*} = -\omega_R + i \omega_I$.
Therefore, we proved that when we perform $e\to -e$, the imaginary part of the QNMs will be the same, while the real part of the QNMs becomes the opposite of itself.
We have also verified this result in our numerics. Fig. \ref{fig:edependence} shows the properties of the QNMs when changing the sign of $e$. Therefore, we can safely focus on positive $e$ to reveal the linear instability. Next, we present the explicit effect of the perturbation charge $e$ on the QNMs, along $\alpha,\, Q,\, \Lambda$, respectively.

\begin{figure}
  \centering
  \includegraphics[width=0.45\textwidth]{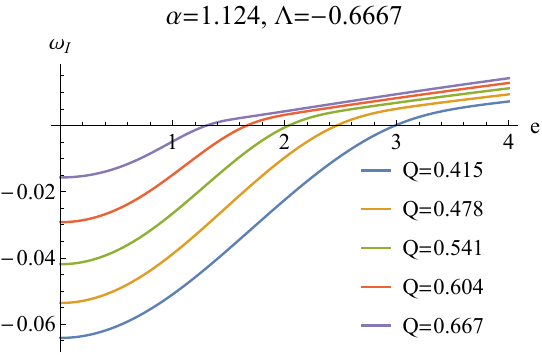}
  \includegraphics[width=0.45\textwidth]{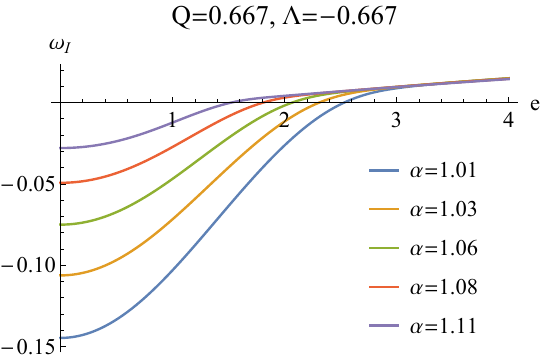}\\
  \includegraphics[width=0.45\textwidth]{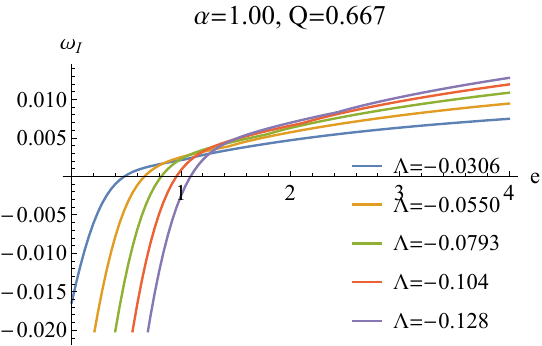}
  \caption{From top to bottom, the QNMs vs $e$ at several different values of $Q,\,\alpha,\,\Lambda$, respectively.}
  \label{fig:qnmvsall}
\end{figure}

The phenomena are qualitatively the same along $\alpha,\, Q,\, \Lambda$, that we demonstrate in Fig. \ref{fig:qnmvsall}. The $\omega_I$ increases with the increase of $e$ for arbitrary parameter. Next, we study the comprehensive instability structure by locating the critical surfaces separating the stable regions and unstable regions.

\subsection{Stable Region}

\begin{figure}[htbp]
  \centering
  \includegraphics[width = 0.7\textwidth]{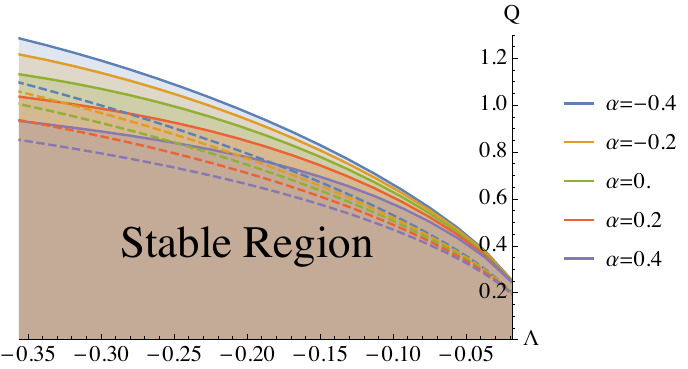}
  \caption{The shaded regions are the stable regions at $e=1,\, l=0$, where each solid line corresponds to different values of $\alpha$ marked by the plot legends. The dashed lines are the critical lines for $e=1.263$, with $\alpha$ specified by the plot legends with the same color.}
  \label{fig:stableregion}
\end{figure}

In order to determine the linear instability structure, we need to locate the critical surfaces on which  $\omega_I$ vanishes. By searching the parameter space subject to the allowed region, we show a comprehensive stable region in Fig. \ref{fig:stableregion} when $e=1,\,l=0$. The linear instability structure in parameter space $(\alpha, \, Q,\, \Lambda)$ are much simpler than that of the dS case \cite{Liu2020}.  Fig. \ref{fig:stableregion} suggests that system becomes more unstable when increasing $Q$ or $\alpha$, while increasing $\Lambda$ will make the system more stable.
Comparing the solid curves $(e=1)$ and the dashed curves $(e=1.263)$ in Fig. \ref{fig:stableregion}, we  find that the stable region indeed shrinks when increasing $e$.
These are consistent with the results from previous subsections.

In the unstable region, we find that the real part of the fundamental modes frequencies of the massless charged scalar perturbation always satisfy the superradiance condition,
\begin{equation}\label{eq:supperradiance}
  \frac{eQ}{r_+} > \omega_R.
\end{equation}
In the stable region, the fundamental modes violate the above condition.
  This suggests that the linear instability from the QNM analysis is triggered by the superradiance instability. We list the QNMs at several critical points of the linear stability/instability transitions in Table \ref{tab:my-table}, from which we can find that $eQ/r_+$ matches perfectly with the real part of the QNMs.
  {Remind that in the dS case \cite{Liu2020}, the superradiance condition is the necessary but not sufficient condition for linear instability.}

\begin{table}[]
  \centering
  \begin{tabular}{|c|c|c|c|}
    \hline
    $\alpha$ & $\Lambda$  & $\frac{eQ}{r_+}$ & $\omega$                                             \\ \hline
    $-0.4$   & ${-}0.337501$ & ${1.25639}$        & ${1.25639} +3.56078\times 10^{-7} i$    \\ \hline
    $-0.4$   & ${-}0.356251$ & ${1.28588}$        & ${1.28588} +8.14388\times 10^{-8} i$  \\ \hline
    $-0.2$   & ${-}0.337501$ & ${1.19356}$        & ${1.19355} +6.51663\times 10^{-7} i$   \\ \hline
    $-0.2$   & ${-}0.356251$ & ${1.21722}$        & ${1.21722} +4.84345\times 10^{-7} i$   \\ \hline
    $0.  $   & ${-}0.337501$ & ${1.11398}$        & ${1.11398} -1.03083\times 10^{-6} i$   \\ \hline
    $0.  $   & ${-}0.356251$ & ${1.13248}$        & ${1.13248} +1.60832\times 10^{-7} i$   \\ \hline
    $0.2 $   & ${-}0.337501$ & ${1.02122}$        & ${1.02122} +7.20995\times 10^{-7} i$   \\ \hline
    $0.2 $   & ${-}0.356251$ & ${1.03673}$        & ${1.03673} -3.68094\times 10^{-7} i$   \\ \hline
    $0.4 $   & ${-}0.337501$ & ${0.919189}$       & ${0.919191} -5.03797\times 10^{-7} i$  \\ \hline
    $0.4 $   & ${-}0.356251$ & ${0.933379}$       & ${0.933378} +1.42506\times 10^{-7} i$  \\ \hline
  \end{tabular}%
  \caption{The fundamental modes at several critical points of the stability-instability transitions corresponding to Fig. \ref{fig:stableregion}.}
  \label{tab:my-table}
\end{table}

\subsection{Time Integral}

In this subsection, we directly implement the time-evolution, i.e., the time integral, of the perturbation field subject to \eqref{eq:main1}. This analysis is important since it can reveal the linear instability structure  in a more transparent manner. Meanwhile, it can work as a double check of the frequency analysis.

On the event horizon, we set the ingoing boundary condition. On the AdS boundary, we set $\phi(v,z=0)=0$ {as required by the asymptotic AdS geometry}. We discretize the spatial direction $z$ with Gauss-Lobatto collocation; while on time direction $v$, we march with the {fourth} order Runge-Kutta method, which has been widely adopted in time evolution problem. Given certain initial profile of the scalar perturbation $\phi$, the time-evolution can be obtained.

\begin{figure}[htbp]
  \centering
  \includegraphics[width = 0.65\textwidth]{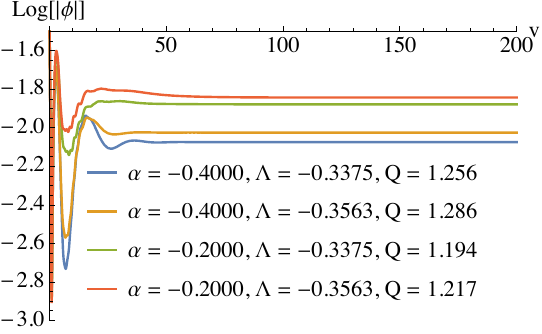}
  \caption{The time integral of the perturbation {$\phi(z=0.997,v)$} at different values of $\alpha, \Lambda, Q$ in the critical region. {The initial profile of $\phi$ is $\phi(z,v=0)=e^{-5{(z-0.5)^2}}$. The time step is $\Delta v = 0.0025$, and we evolve the perturbation to $v=200$.}}
  \label{fig:timeint}
\end{figure}

We show {the time evolution of the $\phi$ near critical lines} in Fig. \ref{fig:timeint}, where we can see that the $\log |\phi|$ quickly becomes linear with $t$. It  seems like all curves are flat, but they have very small slope, as can be  seen from Table. \ref{tab:my-table}. {The small slope (either positive or negative) is  expected since the system is near the critical line for the linear stability/instability transition.}

\begin{figure}[htbp]
  \centering
  \includegraphics[width = 0.6\textwidth]{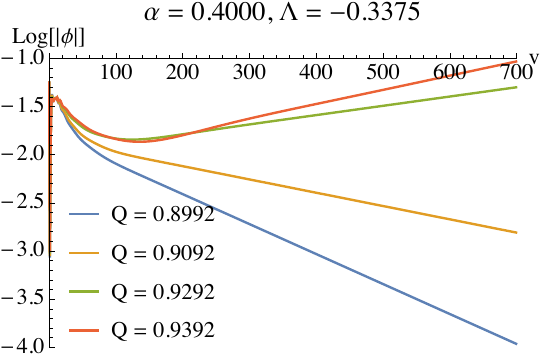}
  \caption{The time integral of the perturbation {$\phi(z=0.997,v)$} at different values of $\alpha, \Lambda, Q$ away from the critical region. {The initial profile of $\phi$ is $\phi(z,v=0)=e^{-5{(z-0.5)^2}}$. The time step is $\Delta v = 0.01$, and we evolve the perturbation to $v=700$.}}
  \label{fig:timeintaway}
\end{figure}

We  worked out  $\partial_t \log |\phi|$ at $t=1000$, which is large enough to obtain stable slopes, i.e., the imaginary part of the dominant modes. Apparently, the slopes should match the results from the frequency analysis. In our numerics we find that they are in good match indeed. For example, the slope for $\alpha = -0.2000, \Lambda= {-}0.3375, Q = 1.194$ is $6.50994\times 10^{-7}$, where the $\omega_I = 6.51663\times 10^{-7}$, as we can see from the third row in Table. \ref{tab:my-table}. These are results near the critical lines, where the numerical precision is hard to control since the $\omega_I$ tends to vanish in critical regions. For regions away from the critical region, the time integral results match much better with those of the frequency analysis.
We show some examples in Fig. \ref{fig:timeintaway}. For the unstable modes, the slopes are positive; while for stable modes, the slopes are negative.
The stabilized slopes do not depend on the specific initial profiles and the locations where we sample the scalar perturbations $\phi$. The above results show that our frequency analysis is robust.

\section{Discussion}\label{sec:discussion}

We studied the linear instability of the {charged massless scalar field in} regularized 4D charged EGB model with asymptotic AdS boundary by examining the QNMs. The detailed linear instability structure of the model was studied by the QNMs vs system parameters $(\alpha, \, Q,\, \Lambda)$ and the scalar charge $e$. We find that the system is unstable against the charged massless scalar when increasing $\alpha$ and $Q$, or {increasing} $\Lambda$. These {phenomena} have been explained intuitively from the viewpoint of the effective potential.
Also, we find that the system is more unstable for larger perturbation charge $e$ and smaller values of $l$. Moreover, the superradiance condition starts to be satisfied across the critical line where linear instability occurs. Time evolution of the perturbation field matches perfectly with the results from the frequency analysis.

Finally, we point out several topics worthy of further study. First, it would be interesting to explore the linear instability structure of the present theory with a massive perturbation. In addition to the scalar perturbation, it is also desirable to reveal the linear instability for tensor perturbations and Dirac fields. {Moreover, the instability of the background metric is also worthy of further studies.}

\section*{Acknowledgments}

We thank Peng-Cheng Li, Minyong Guo for helpful discussions. Peng Liu would like to thank Yun-Ha Zha for her kind encouragement during this work. Peng Liu is supported by the Natural Science Foundation of China under Grant No. 11847055, 11905083. Chao Niu is supported by the Natural Science Foundation of China under Grant No. 11805083. C. Y. Zhang is supported by Natural Science Foundation of China under Grant No. 11947067 and 12005077.

\begin{spacing}{1.2}

\end{spacing}

\end{document}